\documentclass[runningheads]{llncs}
\usepackage[T1]{fontenc}
\usepackage{tabularx}
\usepackage{tikz}
\usetikzlibrary{shapes}
\usetikzlibrary{positioning}
\usetikzlibrary{shapes.geometric}
\usetikzlibrary{arrows.meta}
\usepackage{url}
\usepackage{graphicx}
\begin{document}
	\title{Wearable Data Profiling for NDN-Based Smart Healthcare: Performance Comparison with an IP Baseline\\}
	\titlerunning{Wearable Data Profiling for NDN-Based Smart Healthcare}
	\author{
		Mohammad R. Shakournia\inst{1}\and
		Nasser Yazdani\inst{1}\and
		Carlos T. Calafate\inst{2}
	}
	\authorrunning{M. R. Shakournia et al.}
	\institute{
		Department of Electrical and Computer Engineering, University of Tehran,\\ Tehran, Iran\\
		\email{\{shakournia,yazdani\}@ut.ac.ir}
		\and
		Department of Computer Engineering, Universitat Politècnica de València, Valencia, Spain\\
		\email{calafate@disca.upv.es}
	}
	\maketitle              % typeset the header of the contribution
	\begin{abstract}
		Smart healthcare IoT requires low-latency and reliable delivery of patient data, yet IP-based communication remains host-centric and offers limited support for mobility, caching, and request aggregation. This paper evaluates a smartphone-centric Named Data Networking (NDN) approach in which a patient’s smartphone aggregates sensor streams and publishes a stable “latest profile” under a consistent name, enabling effective in-network caching. Using ndnSIM simulations on a realistic medical IoT topology, we compare NDN and IP under identical traffic and link conditions. Results show that per-sample naming provides little benefit, but smartphone aggregation with stable naming allows NDN caching to reduce delay after cache warm-up and achieve significantly higher delivery ratios under congestion compared to IP. 
		
	\end{abstract}
	\section{Introduction}
	
	The growing use of wearable sensors, home medical devices, and mobile health (mHealth) applications is moving healthcare from occasional, hospital-centered visits to continuous, patient-centered monitoring. In these systems, body-worn and ambient sensors measure vital signs and other physiological parameters and send them, often over short-range and multi-hop wireless links, to a nearby gateway such as a smartphone, which must then forward the data to healthcare staff with low delay and high reliability \cite{Ko2010WSNHealthcare,Ayatollahitafti2016MultiHopBAN}. Many smart-hospital and remote-monitoring platforms already combine IoT, edge, and cloud components to provide such continuous services \cite{Wang2020SecureHealthcareNDNEdge}. However, most of these platforms are built on the traditional TCP/IP protocol stack and therefore inherit its known limits in terms of scalability, mobility support, and efficient, secure data delivery in dense and dynamic IoT environments \cite{Islam2025LeadingSmartEnvironmentsNDN,Azamuddin2023EmergingNDN}. Beyond performance concerns, the underlying networking substrate can also influence how efficiently data and resources are shared across stakeholders, i.e., whether the network itself can facilitate a form of \emph{network-level collective intelligence} by enabling reuse and local sharing of previously retrieved medical information.
	
	The traditional Internet architecture is host-centric: routers forward packets toward IP addresses that identify endpoints rather than the data itself. This model is not well suited for data-intensive smart environments and IoT systems, where applications are mainly interested in named content (for example, the latest ECG of a given patient) instead of a specific server. Recent surveys on NDN in smart environments and IoT report several limitations of the IP model in this setting, including complex address management, limited mobility support, and lack of pervasive in-network caching, which are especially problematic in dense and dynamic IoT healthcare environments. This paper focuses on smartphone-centric IoT healthcare and demonstrates how NDN can overcome these challenges in such settings by offloading much of the traffic management to in-network caches and enabling faster, more efficient data retrieval. In this sense, repeated clinical queries (e.g., by a physician, a clinic system, or an authorized service) can benefit from data that the network has already fetched, effectively turning the network into a shared pool of reusable content. By eliminating the need for traditional IP-based address resolution and supporting mobility, NDN proves to be a better fit for healthcare IoT systems where real-time data from wearable sensors is crucial \cite{Islam2025LeadingSmartEnvironmentsNDN,Shah2022HierarchicalNamingNDNIoT,Amadeo2016ICNIoT,Azamuddin2023EmergingNDN}. These studies argue that simply running application-layer protocols such as MQTT or CoAP over IP cannot fully solve these structural problems; instead, the network layer itself must evolve toward a data-centric design.
	
	Named Data Networking has been proposed as a prominent information-centric networking (ICN) architecture to address these limitations \cite{Zhang2014NDN}. In NDN, applications request content by hierarchical names rather than by host addresses; routers forward \emph{Interest} packets according to name prefixes and return \emph{Data} packets that can be cached at any hop. This leads to a stateful forwarding plane, based on Pending Interest Tables (PITs), Forwarding Information Bases (FIBs), and Content Stores (CSs), which natively supports in-network caching, multipath retrieval, and efficient multicast, while data-centric security is achieved by cryptographically binding each Data packet to its producer, independent of the delivery path \cite{Zhang2014NDN}. As a result, the distributed CS across routers can be viewed as a continuously updated, network-wide knowledge pool: once data is retrieved, it becomes locally available for subsequent consumers without requiring them to know the original producer. Moreover, the cooperative behavior of PIT aggregation and CS hits enables decentralized coordination of data retrieval, allowing the network as a whole to efficiently satisfy repeated requests without a central controller. Surveys on NDN for smart environments and on hierarchical naming for NDN-IoT conclude that these properties align well with the requirements of smart cities, buildings, and health spaces, but also emphasize open issues in naming scalability, resource constraints, and security \cite{Islam2025LeadingSmartEnvironmentsNDN,Shah2022HierarchicalNamingNDNIoT}.
	
	Recent work suggests NDN can improve performance and security in healthcare IoT \cite{Wang2020SecureHealthcareNDNEdge,Dudeja2021EfficientNDNIoTEdge,Dudeja2022SecurePervasiveHealthcare,Islam2025NDNEdgeCloudHealthcare}, but many studies focus on architecture/security/caching strategies rather than a clean baseline comparison against IP under matched conditions \cite{Azamuddin2023EmergingNDN,Islam2025LeadingSmartEnvironmentsNDN,Shah2022HierarchicalNamingNDNIoT,Amadeo2016ICNIoT}. This paper isolates the effect of NDN’s caching and request aggregation in a realistic smartphone-centric healthcare topology and compares delay and delivery reliability to IP.
	
	This work focuses on an underexplored but highly relevant scenario in healthcare IoT, namely a smartphone-centric system, where the patient’s smartphone serves as the data aggregation point for multiple wearable sensors, and acts as the NDN data producer for the patient’s health profile. While the literature has demonstrated the potential of NDN for healthcare, few studies have explicitly tested NDN in small-scale, smartphone-centric topologies that simulate real-world, patient-centric monitoring systems. To address this limitation, we present a comprehensive comparison of NDN and IP in a smartphone-centric healthcare scenario, evaluating their performance under identical experimental conditions. Our study demonstrates that NDN can improve real-time data delivery by exploiting data-centric communication and in-network caching while requiring only minimal modifications to existing infrastructure. Experimental results further indicate that this architecture reduces communication latency and better accommodates user mobility, highlighting the potential of NDN as a practical networking solution for future smart healthcare systems.
	
	The rest of this paper is organized as follows. Section~II reviews background on NDN and summarizes related work on NDN-based healthcare and smart environments. Section~III presents the proposed system model, including node roles, naming scheme, and traffic types. Section~IV describes the simulation setup and performance metrics. Section~V discusses the numerical results comparing NDN and IP. Finally, Section~VI concludes the paper and outlines directions for future work.
	
	\section{Related Work}
	
	Early IP-based BAN/WSN healthcare systems forward sensor data via gateways over TCP/IP (often 6LoWPAN), but the host-centric model offers limited support for mobility, in-network caching, and data-centric security.\cite{Ko2010WSNHealthcare,Ayatollahitafti2016MultiHopBAN}.
	
	Information-centric networking, and in particular NDN, has been proposed to better match the data-oriented nature of IoT and smart environments. Amadeo \textit{et al.} review ICN for the Internet of Things and argue that name-based access, in-network caching, and data-centric security can address many of the scalability and mobility issues that arise under IP \cite{Amadeo2016ICNIoT}. Zhang \textit{et al.} present the NDN architecture, in which consumers issue Interests for named content and Data packets can be served from any node that holds a copy, whether the original producer or an intermediate cache \cite{Zhang2014NDN}. Recent surveys extend this perspective to smart environments and NDN-IoT deployments. Islam \textit{et al.} show that NDN’s hierarchical naming, stateful forwarding, and built-in caching align well with the needs of smart cities, buildings, and health spaces, but also highlight practical barriers such as interoperability with IP and limited testbeds \cite{Islam2025LeadingSmartEnvironmentsNDN}. Shah \textit{et al.} focus on hierarchical and hybrid naming schemes for NDN-based IoT, including e-health scenarios, and discuss how naming choices interact with access control, content protection, and the resource constraints of IoT devices \cite{Shah2022HierarchicalNamingNDNIoT}. Azamuddin \textit{et al.} provide a broad overview of NDN architecture, applications, and open research problems, further motivating its use in emerging IoT and edge-computing settings \cite{Azamuddin2023EmergingNDN}. On the caching side, Jafari \textit{et al.} propose ECLAT-LRU, a predictive caching policy that improves NDN cache performance \cite{11309195}.
	
	On top of this architectural groundwork, several NDN-based frameworks for healthcare and related IoT domains have been proposed. Wang and Cai introduce SHNIE, a secure healthcare monitoring framework that integrates clustered IoT devices with an NDN-based edge cloud \cite{Wang2020SecureHealthcareNDNEdge}. SHNIE uses hash-based ciphertext names, encryption, and signatures to protect medical data and exploits request aggregation and in-network caching to reduce communication latency and message cost compared with an IP-based smart hospital system. Dudeja \textit{et al.} design an NDN-based data communication framework for IoT--edge ecosystems in which edge nodes act as cluster heads and assign priorities to periodic sensor data; these priorities are then mapped to different caching strategies (e.g., Leave-Copy-Everywhere versus Leave-Copy-Down) in the NDN core \cite{Dudeja2021EfficientNDNIoTEdge}. In a follow-up work, they propose a secure and pervasive NDN communication framework for connected healthcare that supports both consumer-initiated and producer-initiated data dissemination and evaluate its performance under various network conditions \cite{Dudeja2022SecurePervasiveHealthcare}. Demir and Manap implement a Named Data Networking of Things (NDNoT) stack on the Contiki-NG operating system and demonstrate the feasibility of NDN-based medical data retrieval on constrained IoT devices using Cooja simulations \cite{Demir2022MedicalDataRetrieval}. Islam \textit{et al.} outline an NDN-aided edge--cloud architecture for continuous smart healthcare, where NDN is used at the edge and an NDN edge gateway bridges local NDN domains to IP-based cloud services such as AWS IoT \cite{Islam2025NDNEdgeCloudHealthcare}.
	
	These efforts collectively illustrate the potential benefits of NDN in healthcare and IoT-edge settings, including improved content delivery efficiency, flexible data dissemination patterns, and stronger data-centric security. However, most existing works either focus on security mechanisms and naming schemes, on architectural integration with edge and cloud platforms, or on NDN-only performance tuning. SHNIE does include an analytical and simulation-based comparison with an IP smart-hospital baseline, but its design centers on a relatively complex edge-cloud security framework rather than on a minimal end-to-end communication model \cite{Wang2020SecureHealthcareNDNEdge}. Other proposals evaluate NDN-based healthcare or IoT frameworks in isolation, without constructing a matching IP reference system over the same topology and traffic patterns \cite{Dudeja2021EfficientNDNIoTEdge,Dudeja2022SecurePervasiveHealthcare,Demir2022MedicalDataRetrieval,Islam2025NDNEdgeCloudHealthcare}. Despite recent advances, controlled side-by-side comparisons between NDN and IP under identical experimental conditions remain limited, particularly in smartphone-centric medical IoT scenarios where the patient's smartphone serves as both the edge device and the primary data producer. In such environments, remote caregivers retrieve up-to-date patient information directly by name, making it possible to evaluate the practical benefits and limitations of NDN relative to conventional IP networking.

	\section{System Model and Proposed Architecture}
	
	We study the use of NDN in a smart healthcare system where patients are equipped with wearable IoT sensors. Our goal is to evaluate whether NDN can provide communication performance comparable to, or better than, a traditional IP-based solution when delivering medical data from patients to remote caregivers. In addition to end-to-end performance, we are interested in whether NDN’s data-centric mechanisms enable efficient reuse of previously retrieved health information across multiple authorized stakeholders, effectively leveraging the network as a shared substrate for data access.
	
	\subsection{Overall Scenario and Node Types}
	
	We consider a hospital or clinic scenario in which each patient is continuously monitored by a set of body-worn sensors. These sensors periodically generate physiological measurements such as heart rate, SpO$_2$, and blood pressure. A patient’s smartphone acts as a local gateway and edge device that collects sensor data and makes it available to remote caregivers.
	
	In our evaluation, we model six patients, each with one smartphone. The smartphones and caregivers are interconnected by a fixed multi-hop network. On the caregiver side, we consider six monitoring consumers representing clinics, physicians, or hospital monitoring stations. This multi-consumer setting naturally captures repeated access to the same patient state from different monitoring points, which is precisely where in-network caching can provide network-wide reuse by serving subsequent requests from nearby nodes. Between smartphones and consumers, the network includes nine routers (e.g., access routers and intermediate/core routers), resulting in a 21-node topology. The same topology, node roles, and traffic patterns are used in both NDN and IP scenarios (Section~IV will detail link capacities, propagation delays, and traffic parameters).
	
	The medical traffic of interest consists primarily of periodic monitoring flows (vital signs) and time-critical updates (e.g., alarms or abrupt changes). We focus on end-to-end delay and packet loss as the main performance metrics; all security mechanisms (encryption, signatures) are abstracted away and not modeled in the simulations.
	\vspace{-0.05cm}
	\subsection{Baseline NDN Design: Direct Sensor-to-Consumer Communication}
	As an initial baseline, we consider a straightforward NDN design in which each body-worn sensor acts as an NDN data producer. Every new measurement generated by the sensor (e.g., once per second, once per minute, or at any other configured sampling interval) is published as a distinct NDN Data packet with a unique name that encodes the generation time. In this model, clinics, doctors, or hospital servers act as NDN consumers and retrieve samples directly from the sensors by issuing Interests for these time-stamped names over a multi-hop topology, similar to IP-based body area network designs where sensors forward data to a sink or gateway \cite{Ko2010WSNHealthcare,Ayatollahitafti2016MultiHopBAN}. For such \emph{time-critical} flows, where each sample is unique and must be delivered as soon as it is produced, in-network caching and request aggregation offer little benefit, because each measurement is requested exactly once and is not reused by other consumers. As a result, this baseline intentionally represents a setting where the network cannot exploit reuse across stakeholders, since there is no shared, repeatedly requested content to retain in the forwarding plane.
	
	To better isolate this effect, we also conduct a micro-benchmark in a minimal topology consisting of a single producer (modeling a body-worn sensor), one intermediate router, and a single consumer. In this setup, the producer generates 1000 distinct packets, each carrying a unique name (or, in the IP case, a unique sequence identifier), and we transmit the same 1000 packets over both an IP and an NDN stack under identical link settings, measuring the end-to-end request–response delay. Since all packets are strictly non-reusable, this experiment intentionally removes the main performance advantages of NDN, namely in-network caching and request aggregation. Intermediate NDN routers still perform per-packet lookups in CS and PIT, in addition to FIB lookups, whereas the IP baseline only maintains a routing table and forwards packets without per-request state. In our ns-3/ndnSIM implementation, the additional CS/PIT processing overhead is modeled at a logical level and therefore manifests as only a small difference in simulated delay; however, in a real implementation with non-negligible lookup and memory costs, this overhead can become more pronounced and may make NDN slower than a stateless IP forwarder under purely time-critical, non-cacheable traffic. The numerical results of this sanity-check experiment, summarized in Table~\ref{tab:sim-params-results}, confirm that under these conditions NDN does not provide a delay advantage over IP.
	
	\begin{table}[!t]
		\centering
		\caption{Simulation parameters and RTT results for the single-hop baseline scenario}
		\label{tab:sim-params-results}
		
		% ---------- Part (a): Simulation parameters ----------
		\begin{tabular}{ll}
			\multicolumn{2}{c}{\textbf{(a) Simulation parameters}}\\
			\hline
			Parameter      & Value            \\
			\hline
			Link data rate & 10 Mbps          \\
			Link delay     & 1 ms             \\
			Queue size     & 50 packets       \\
			Traffic Size   & 1000               \\
			\hline
		\end{tabular}
		
		\vspace{2mm}
		
		% ---------- Part (b): Experimental results ----------
		\begin{tabular}{lcc}
			\multicolumn{3}{c}{\textbf{(b) RTT statistics at consumer}}\\
			\hline
			Metric             & IP (Request--Response) & NDN (Interest--Data) \\
			\hline
			Samples            & 1000                   & 1000                 \\
			Mean [ms]          & 4.736                  & 5.447                \\
			Median [ms]        & 4.736                  & 5.448                \\
			95th perc. [ms]    & 4.736                  & 5.448                \\
			99th perc. [ms]    & 4.736                  & 5.448                \\
			\hline
		\end{tabular}
	\end{table}

	\subsection{Proposed Smartphone-Centric NDN Architecture}
	
	To enable NDN to exploit its strengths in this setting, we propose a smartphone-centric architecture in which the patient’s smartphone acts as an edge device and the primary NDN data producer.
	
	\begin{figure*}[t]
		\centering
		\includegraphics[width=0.8\textwidth]{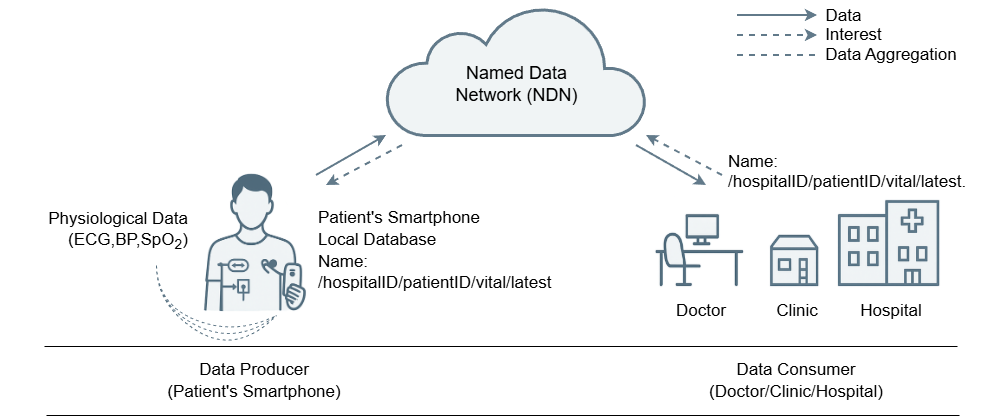}
		\caption{System architecture of the proposed NDN-enabled IoHT environment. The patient’s smartphone acts as the data producer, aggregating sensor readings and serving named profile data to healthcare consumers (clinics, physicians, hospitals) through the Named Data Network.}
		\label{fig:system_architecture}
	\end{figure*}
	
	\subsubsection{Body-Worn Sensors and Smartphone Edge Device}
	
	In the proposed architecture, each patient wears multiple body-mounted sensors that measure physiological parameters such as heart rate, SpO$_2$, and blood pressure. These sensors communicate wirelessly with the patient's smartphone, which acts as a local edge device and data aggregator. In line with current commercial practice, we assume that body-worn sensors use Bluetooth Low Energy (BLE) to connect to the smartphone, which plays the role of the BLE central device and periodically collects measurements from multiple BLE peripherals \cite{Zhang2014BLEHealthcare,Chang2023BluetoothHealthcare}. This reflects how many consumer-grade wearable healthcare and fitness devices (e.g., chest straps, smartwatches, and glucose monitors) deliver physiological data to mobile phones in real deployments.
	
	In our system model, the details of the BLE link (pairing, GATT services, etc.) are abstracted into a single logical low-power link from the sensor set to the smartphone. We focus on the end-to-end delivery of aggregated medical data from the smartphone to remote caregivers over either NDN or IP.
	
	\subsubsection{NDN Naming and Personal Health Profile}
	
	The smartphone continuously aggregates the incoming sensor readings into a \emph{personal health profile} that is updated whenever new data arrive. Instead of publishing every raw sample as a separate NDN object, the smartphone exposes a single named NDN Data object that always represents the latest state of the patient’s vital signs. For a patient with identifier \texttt{patientID}, the current profile may be published under a stable name of the form
	\[
	/\textit{hospital}/\textit{patientID}/\textit{vital}/\textit{latest}.
	\]
	The smartphone therefore acts as the sole NDN producer for this profile, while clinics, doctors, and hospital servers act as NDN consumers that issue Interests for this name whenever they need the most recent health status.
	
	Because all monitoring systems use the same content name to retrieve the patient’s latest profile, intermediate NDN routers and edge nodes can cache the Data packet and satisfy subsequent Interests from nearby caches. This is particularly relevant for smart-healthcare monitoring, where multiple monitoring points (e.g., ward dashboards, nurse stations, physician terminals) repeatedly request the same patient’s current status. In this way, the distributed caches collectively act as a shared pool of recently retrieved patient profiles, enabling consumers to obtain the closest available copy without needing to target a particular host. In our evaluation (Section~IV), we consider several such monitoring consumers that periodically retrieve the profile of each patient; once a profile is fetched the first time, subsequent requests can be served from caches, and NDN’s in-network caching begins to provide a clear benefit.
	
	\subsection{Consistency and Freshness Semantics}
	
	A key requirement in this architecture is that caregivers must receive up-to-date information. NDN provides built-in mechanisms to express freshness requirements. Each Data packet includes a \texttt{FreshnessPeriod} indicating how long cached copies should be considered valid, and consumers may set the \texttt{MustBeFresh} selector in their Interests to request only fresh Data.
	
	In our design, we align the \texttt{FreshnessPeriod} of the health-profile Data with the sensor update period. As a result, cached copies of the profile automatically expire in routers roughly when new data are produced, and Interests marked as \texttt{MustBeFresh} are forwarded toward the smartphone if cached content has become stale. This ensures that doctors and monitoring systems obtain a consistent and up-to-date view of the patient’s vital signs while still benefiting from in-network caching between updates. Put differently, freshness semantics help keep the shared cached pool reliable by preventing stale profiles from being served beyond their validity window.
	
	A high-level representation of the proposed architecture is shown in Fig.~2. Wearable sensors attached to each patient send data to the patient’s smartphone, which acts as an edge device and NDN producer. A set of intermediate routers connects the smartphones to multiple monitoring consumers (clinics, doctors, and hospital servers). NDN runs from the edge devices to the consumers, using hierarchical names and in-network caching, while the specific low-power protocol between sensors and the smartphone is abstracted away in the system model.
	
	\subsection{IP Reference Architecture}
	
	For a fair comparison, we also construct an IP reference architecture that uses the same physical topology, node roles, and traffic patterns as the NDN architecture. In the IP baseline, each patient’s smartphone still collects sensor data via BLE and maintains an internal health profile, but communication with caregivers follows the conventional host-centric model.
	
	Specifically, the smartphone transmits monitoring data to caregiver nodes using UDP over IP. Intermediate routers maintain IP routing tables and forward packets based on destination IP addresses. There is no in-network content caching in the network: routers do not store application-layer data, and each monitoring request must be served by the smartphone or a backend server every time. Consequently, the network does not accumulate reusable content state on the path, and repeated monitoring requests cannot benefit from previously retrieved data at intermediate nodes. No NDN-specific state (such as PIT or CS) is present. This IP baseline allows us to quantify the impact of NDN’s name-based forwarding and in-network caching on end-to-end delay and packet loss under identical network and traffic conditions.
	\section{Simulation Setup and Metrics}
	
	We compare the proposed smartphone-centric NDN architecture with an IP baseline in \texttt{ns-3}, using the standard Internet stack for IP and ndnSIM for NDN. Both scenarios share the same physical topology, node roles, and traffic patterns; only the network layer (IP vs.\ NDN with name-based forwarding and caching) differs.
	
	\subsection{Topology and Traffic Overview}
	
	We use the 21-node topology from Section~III: six monitoring consumers, six patient-side producer/server nodes (smartphones), and nine routers connected through a constrained core. All consumer--producer traffic traverses the same core bottleneck, enabling controlled congestion and loss experiments.
	
	Each consumer is associated with one patient. In IP, consumers send UDP traffic to the corresponding patient-side server; in NDN, they issue Interests to the corresponding producer. Applications run over the same time interval in both cases, and packet/payload sizes are matched for fairness. Table~\ref{tab:sim-params} summarizes the main simulation parameters.

	\begin{figure}[!t]
		\centering
		\resizebox{0.65\columnwidth}{!}{% adjust 0.9 to 1.0 if you want full width
			
			\begin{tikzpicture}[
				>=Latex,
				node distance=1.1cm and 1.5cm,
				router/.style={rectangle, draw, rounded corners, minimum width=1.1cm, minimum height=0.7cm, thick, fill=blue!5},
				host/.style={circle, draw, minimum size=0.6cm, thick, fill=green!5},
				phone/.style={rounded rectangle, draw, minimum width=0.9cm, minimum height=0.7cm, thick, fill=orange!10},
				link/.style={thick},
				font=\small
				]
				
				% ----------------------
				% Left side: consumers
				% ----------------------
				
				\node[host] (c1) {C1};
				\node[host, below=of c1] (c2) {C2};
				\node[host, below=of c2] (c3) {C3};
				
				\node[host, right=6cm of c1] (c4) {C4};
				\node[host, below=of c4] (c5) {C5};
				\node[host, below=of c5] (c6) {C6};
				
				% ----------------------
				% Edge routers (monitoring side)
				% ----------------------
				
				\node[router, right=0.5cm of c2] (re1) {R1};
				\node[router, left=0.5cm of c5] (re2) {R2};
				
				% Connect consumers to edge routers
				\foreach \i in {1,2,3}{
					\draw[link] (c\i) -- (re1);
				}
				\foreach \i in {4,5,6}{
					\draw[link] (c\i) -- (re2);
				}
				
				% ----------------------
				% Core router
				% ----------------------
				
				\node[router, right=1cm of re1, yshift=-1.5cm] (rc) {R3};
				
				\draw[link] (re1) -- (rc);
				\draw[link] (re2) -- (rc);
				
				% ----------------------
				% Patient-side routers
				% ----------------------
				
				% Three upper patient routers
				\node[router, below=of rc, xshift=-3.7cm] (rp1) {ER1};
				\node[router, below=of rc, xshift=-2.5cm] (rp2) {ER2};
				\node[router, below=of rc, xshift=-1.3cm] (rp3) {ER3};
				
				% Three lower patient routers
				\node[router, below=of rc, xshift=0.7cm] (rp4) {ER4};
				\node[router, below=of rc, xshift=1.9cm] (rp5) {ER5};
				\node[router, below=of rc, xshift=3.1cm] (rp6) {ER6};
				
				% Connect core to patient routers
				\foreach \x in {rp1,rp2,rp3,rp4,rp5,rp6}{
					\draw[link] (rc) -- (\x);
				}
				
				% ----------------------
				% Smartphones / producers
				% ----------------------
				
				\node[phone, below=1.2cm of rp1] (p1) {P1};
				\node[phone, below=1.2cm of rp2] (p2) {P2};
				\node[phone, below=1.2cm of rp3] (p3) {P3};
				\node[phone, below=1.2cm of rp4] (p4) {P4};
				\node[phone, below=1.2cm of rp5] (p5) {P5};
				\node[phone, below=1.2cm of rp6] (p6) {P6};
				
				% Connect patient routers to phones
				\foreach \i in {1,2,3,4,5,6}{
					\draw[link] (rp\i) -- (p\i);
				}
				
				\node[below=4cm of rc] {\footnotesize Smartphones / NDN producers};
				\node[above=2.8cm of rc] {\footnotesize Clinic/Doctor/Hospital(NDN consumers)  };

				% Optional labels for domains
				% \node[draw=none, fit=(c1)(c6)(re2), inner sep=0.3cm, label=above:{\footnotesize Monitoring domain}] {};
				% \node[draw=none, fit=(rp1)(rp6)(p6), inner sep=0.3cm, label=below:{\footnotesize Patient domain}] {};
				
			\end{tikzpicture}
		}
		\caption{21-node topology with six monitoring consumers, nine routers (two edge routers, one core, six patient-side routers), and six patient smartphones acting as NDN producers.}
		\label{fig:topology}
	\end{figure}
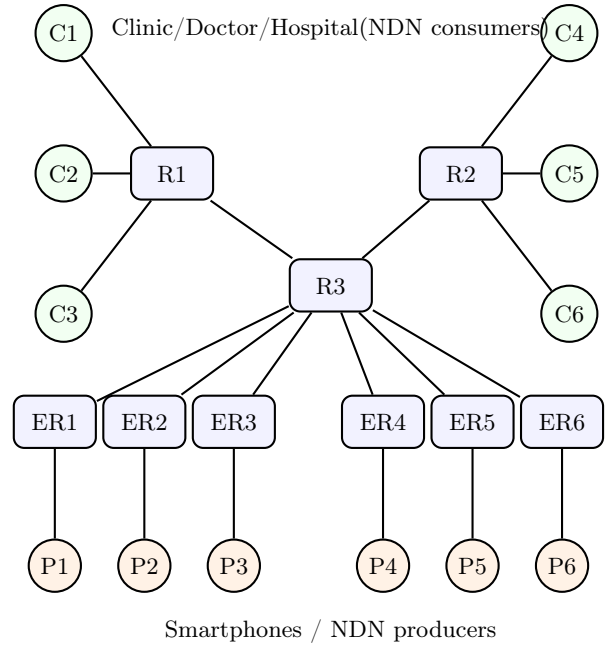

	\begin{table}
		\caption{Comparison of IP and NDN Simulation Parameters}
		\label{tab:sim-params}
		\centering
		\renewcommand{\arraystretch}{1.15}
		\begin{tabular}{lll}
			\hline
			\textbf{Parameter} & \textbf{IP} & \textbf{NDN} \\
			\hline
			Total traffic & 1M packets & 1M Interests/Data \\
			Application model & OnOff & ConsumerCbr \\
			Payload size & 900 bytes & 900 bytes \\
			Traffic rate per consumer & 72 Kbps & 10 Interests/s \\
			Forwarding strategy & Default & Best-route / Multicast \\
			Interest lifetime & -- & 2 s \\
			CS size & -- & 1000 Data packets \\
			Link bandwidth & 10 Mbps & 10 Mbps \\
			Link delay & 2 ms & 2 ms \\
			Queue size & 1000 & 1000 \\
			\hline
		\end{tabular}
	\end{table}
	
	Unless otherwise stated, NDN uses ndnSIM's \texttt{MulticastStrategy}; we also tested \texttt{BestRoute}. In this static, tree-like topology, both yield nearly identical RTT and delivery ratio, indicating that the observed gains mainly come from in-network caching rather than path diversity.
	
	\subsection{Metrics}
	
	We use two primary performance metrics:
	
	\begin{itemize}
		\item \textbf{Round-trip delay (RTT):}
		Both IP and NDN follow a request--response model. In IP, RTT is measured from sending a probe packet to receiving its acknowledgment; in NDN, from issuing an Interest to receiving the corresponding Data. We report RTT under normal conditions and under packet loss induced by core-link congestion.
		
		\item \textbf{Delivery ratio:}
		In IP, this is the fraction of probes that receive acknowledgments; in NDN, it is the Interest satisfaction ratio (Data received / Interests sent). This metric captures the impact of congestion, loss, and in-network caching.
	\end{itemize}
	
	Using identical topology, traffic roles, and packet sizes, these metrics enable a fair, side-by-side comparison of IP and NDN performance in the smartphone-centric healthcare scenario.
	\vspace{-0.2cm}
	\section{Results and Discussion}
	
	We compare end-to-end performance of IP and NDN under (i) sensor-level,
	per-sample naming and (ii) a smartphone-centric, profile-based naming
	model that enables in-network caching.
	
	\subsection{Baseline Sensor-Level Traffic}
	
	Table~I compares IP and NDN when each sensor reading is sent as a
	separate packet and each NDN Data packet carries a unique time-stamped
	name. Because every measurement is distinct, caches cannot be reused and
	each Interest must reach the producer; moreover, NDN routers incur extra
	per-packet PIT/CS processing. Consequently, NDN does not outperform IP
	in this sensor-level setting and may show a small processing penalty.
	This suggests that for strictly time-critical traffic with non-reusable
	samples, exposing every measurement as an independent NDN object is
	ineffective.
	
	Based on this observation, the main experiments use a profile-based
	model: the smartphone aggregates sensor readings into a continuously
	updated personal health profile and publishes it under a stable name
	(e.g., \path{/HospitalID/patientID/vital/latest}). As summarized in
	Table~\ref{tab:sim-params}, repeated queries for the same patient map to
	the same name and can benefit from in-network caching.
	
	\subsection{Short-Term Cache Warm-Up Behavior}
	
	Figure~\ref{fig:rolling-mean-results} shows a 10-packet rolling mean RTT
	over the first 2500 packets. In both loss and no-loss runs, NDN exhibits
	a short initial RTT spike corresponding to cache warm-up, when early
	Interests traverse the full multi-hop path to the smartphone.
	
	After a few requests, the profile Data are cached near the monitoring
	stations, and subsequent Interests are satisfied locally, causing NDN
	RTT to drop quickly to a low, nearly constant value. IP shows no such
	warm-up effect: RTT stays roughly constant without loss, but grows and
	oscillates under congestion. These results confirm that NDN gains here
	are driven by caching profiles close to consumers.

	\subsection{Delay Under Packet Loss}
	
	To induce congestion and loss, core links are set to 200~kbps, making the
	core the dominant bottleneck and causing frequent queue drops. As shown
	in the upper plots of Figs.~\ref{fig:rolling-mean-results} and
	\ref{fig:chunked-results}, IP exhibits unstable RTT with pronounced
	oscillations and values often exceeding $10^3$~ms, due to repeated queue
	overflows and retransmissions in the request--response exchange.
	
	NDN remains consistently low, with chunk-averaged RTT on the order of
	$10^2$~ms under the same conditions. Because requests target the same
	named profile, fetched Data are cached on the consumer side of the
	bottleneck, so subsequent Interests are satisfied locally without
	crossing the congested core. As a result, the rolling-mean curve flattens
	after warm-up and Fig.~\ref{fig:chunked-results} shows a narrow RTT band
	throughout the one-million-packet run.
	
	\subsection{Delay Without Packet Loss}
	
\begin{figure}[!t]
		\centering
		\includegraphics[width=0.9\columnwidth]{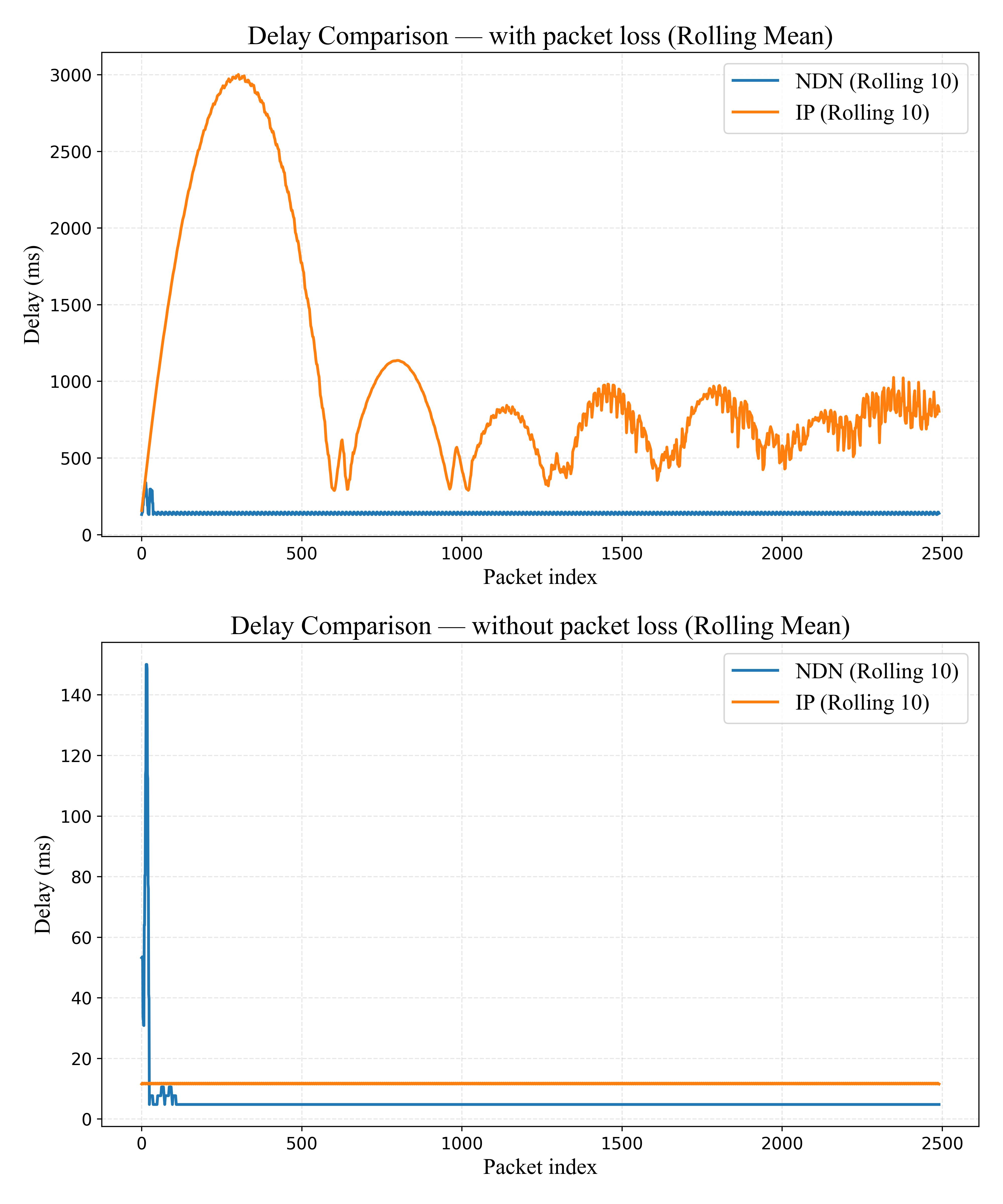}
		\caption{Rolling-mean RTT (window 10) over the first 2500 packets for IP
			and NDN under packet loss (top) and no loss (bottom), highlighting cache
			warm-up in NDN.}
		\label{fig:rolling-mean-results}
	\end{figure}

\begin{figure}[!t]
		\centering
		\includegraphics[width=0.9\columnwidth]{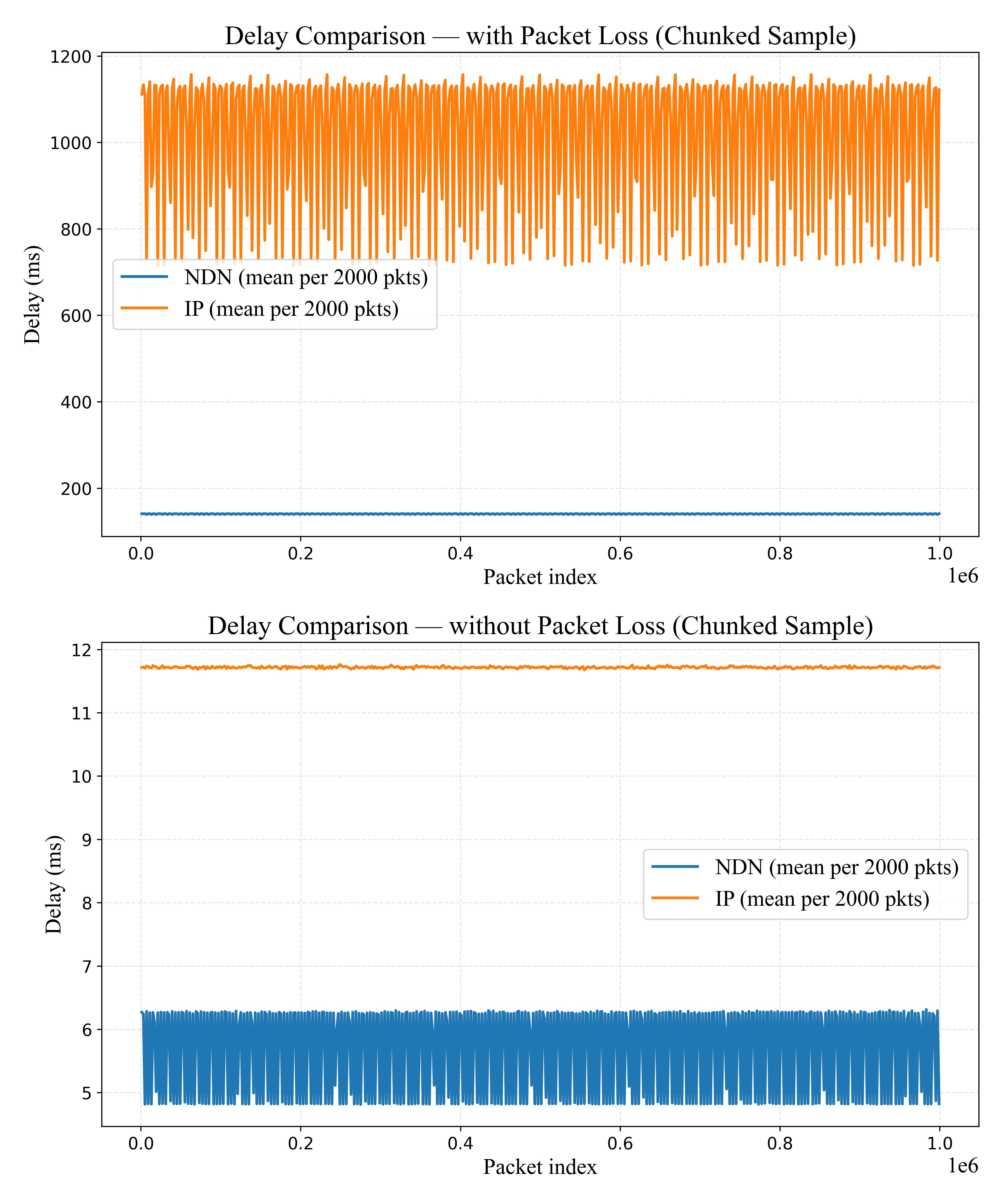}
		\caption{Chunk-averaged RTT (mean per 2000 delivered packets) for IP and
			NDN under packet loss (top) and no loss (bottom) over one million packets.}
		\label{fig:chunked-results}
	\end{figure}

	Without loss, both architectures are stable but differ in absolute RTT.
	In the lower plots of Figs.~\ref{fig:rolling-mean-results} and
	\ref{fig:chunked-results}, IP remains near $10$--$12$~ms, reflecting
	fixed queuing and propagation along the multi-hop path.
	
	NDN converges to about $5$--$6$~ms after a brief cache warm-up. Once the
	profile is cached near consumers, Interests are satisfied over shorter
	effective paths and repeated upstream traversal through the core is
	avoided, yielding persistently lower RTT over the one-million-packet run.
	
	\subsection{Delivery Ratio}
	
	Table~\ref{tab:delivery-ratio} reports end-to-end delivery. In the
	no-loss scenario, both IP and NDN achieve near-perfect delivery. Under
	packet loss on the 200~kbps core, IP delivery drops sharply to about
	$3\%$, whereas NDN remains close to $99\%$, because caching reduces the
	amount of traffic that must traverse the lossy core. This shows that
	NDN caching improves both delay and robustness under congestion.
	
	Overall, NDN is not inherently faster than IP for per-packet,
	time-critical sensor traffic because unique names prevent cache reuse and
	PIT/CS processing adds overhead. In contrast, the smartphone-centric,
	profile-based model maps repeated monitoring to a stable name, enabling
	routers near monitoring stations to serve requests from cache. This
	yields lower and more stable RTT and much higher delivery ratios than IP
	in lossy or resource-limited settings, consistent with continuous remote
	monitoring where caregivers primarily require an up-to-date patient
	profile rather than every raw sample.

	\begin{table}
		\caption{End-to-end delivery ratio for IP and NDN under loss-free and packet-loss conditions.}
		\label{tab:delivery-ratio}
		\centering
		\renewcommand{\arraystretch}{1.1}
		\begin{tabular}{lcc}
			\hline
			Scenario & IP delivery ratio & NDN delivery ratio \\
			\hline
			No-loss & 1.00 & 1.00 \\
			Packet-loss (200 kbps core) & 0.03 & 0.99 \\
			\hline
		\end{tabular}
	\end{table}
	
	\section{Conclusion}
	
	This paper revisited the role of Named Data Networking in smartphone-centric healthcare systems by contrasting a baseline, sensor-level traffic model with a profile-based design in which the patient’s smartphone aggregates measurements into an abstract, continuously updated health profile published under a stable NDN name. The baseline results showed that NDN is not inherently faster than IP for strictly time-critical, per-packet sensor traffic, because unique names limit cache reuse and PIT/CS processing does not translate into lower delay. Under congestion with loss, our NDN approach maintains low, stable RTT and near-unity delivery, whereas IP degrades to second-scale delays and only a few percent delivery; in the loss-free case, NDN halves RTT after a brief cache warm-up. Taken together, the results indicate that NDN's key strength in smartphone-centric healthcare lies in the combination of data-centric naming, smartphone-level profile aggregation, and in-network caching. These mechanisms reduce communication latency and improve resilience under network congestion, making them well suited for continuous remote patient monitoring. In addition, caching and Interest aggregation enable authorized healthcare providers to reuse nearby profile replicas and coordinate data retrieval without relying on a centralized controller, thereby improving scalability and communication efficiency. Future work will focus on evaluating larger mobile topologies, incorporating more realistic wireless channel models and traffic patterns, and investigating the integration of NDN's trust-based security and privacy mechanisms.
	\bibliographystyle{splncs04}
	\bibliography{references}

\end{document}